\documentclass{webofc}

\usepackage[varg]{txfonts}   %
\usepackage{hyperref}
\usepackage{url}
\hypersetup{colorlinks=true,citecolor=blue,urlcolor=blue,linkcolor=blue}

\usepackage{xcolor}
\definecolor{myOrange}{rgb}{1,0.5,0.}
\definecolor{myGreen}{rgb}{0.0,0.6,0.1}

\newcommand{\RAA}          {\ensuremath{R_{\mathrm{AA}}}\xspace}

\newcommand{\nineH}        {$\sqrt{s}~=~0.9$~Te\kern-.1emV\xspace}
\newcommand{\seven}        {$\sqrt{s}~=~7$~Te\kern-.1emV\xspace}
\newcommand{\twoH}         {$\sqrt{s}~=~0.2$~Te\kern-.1emV\xspace}
\newcommand{\twosevensix}  {$\sqrt{s}~=~2.76$~Te\kern-.1emV\xspace}
\newcommand{\five}         {$\sqrt{s}~=~5.02$~Te\kern-.1emV\xspace}
\newcommand{\twosevensixnn}{$\sqrt{s_{\mathrm{NN}}}~=~2.76$~Te\kern-.1emV\xspace}
\newcommand{\fivenn}       {$\sqrt{s_{\mathrm{NN}}}~=~5.02$~Te\kern-.1emV\xspace}

\newcommand{\GeVc}         {Ge\kern-.1emV/$c$\xspace}
\newcommand{\MeVc}         {Me\kern-.1emV/$c$\xspace}
\newcommand{\TeV}          {Te\kern-.1emV\xspace}
\newcommand{\GeV}          {Ge\kern-.1emV\xspace}
\newcommand{\MeV}          {Me\kern-.1emV\xspace}
\newcommand{\GeVmass}      {Ge\kern-.2emV/$c^2$\xspace}
\newcommand{\MeVmass}      {Me\kern-.2emV/$c^2$\xspace}

\newcommand{\qhat}{\ensuremath{\hat{q}}\xspace}
\newcommand{\qhatTcubed}{\ensuremath{\hat{q}/{T}^3}\xspace}

\newcommand{\sqrtsNN}{\ensuremath{\sqrt{s_\mathrm{NN}}}}

\newcommand{\pT}{\ensuremath{p_\mathrm{T}}}

\begin{document}
\title{Bayesian inference and jet quenching}

\author{\firstname{Raymond} \lastname{Ehlers}\inst{1,2}\fnsep\thanks{\email{raymond.ehlers@cern.ch}}
}

\institute{Nuclear Science Division, Lawrence Berkeley National Laboratory, Berkeley, CA 94720, USA
\and
           Department of Physics, University of California, Berkeley, CA 94720, USA 
          }

\abstract{
These proceedings review the application of Bayesian inference to high momentum transfer probes of the quark--gluon plasma (QGP).
Bayesian inference techniques are introduced, highlighting critical components to consider when comparing analyses.
Recent calibrations using hadron observables are described, illustrating the importance of the choice of parametrization.
Additional recent analyses that characterize the impact of the inclusion of jet observables, as well as soft-hard correlations, are reviewed.
Finally, lessons learned from these analyses and important questions for the future are highlighted.
}
\maketitle

\section{Introduction to Bayesian inference}

Jet quenching measurements performed at the Relativistic Heavy Ion Collider (RHIC) and the Large Hadron Collider (LHC) over the last two decades contain a wealth of information about the quark--gluon plasma (QGP). 
These high-momentum transfer processes probe the dynamics and structure of the QGP.
However, multiple model calculations based on different underlying physical formalisms are able to describe individual measurements equally well; such single-observable comparisons provide limited discrimination of different physical mechanisms.
Due to the complex nature of the many-body dynamics of the medium, rigorous and systematic data-model comparison of multiple observables, which probe the same underlying physical processes in significantly different ways, is required to fully interpret the underlying physics.
Bayesian inference provides an ideal tool for such comparisons.
Early Bayesian analyses in heavy-ion collisions have focused on the soft sector~\cite{Bernhard:2019bmu, JETSCAPE:2020shq, Nijs:2020roc}, while more recent developments in the hard sector are described in these proceedings.

Given a model, which parameters are most compatible with experimental measurements?
Bayesian inference is a rigorous method for data-model comparison, providing a systematic approach to answering this question.
This comparison is performed via Bayes' theorem,

\begin{equation}
   P(\vec{\theta} | \vec{x}) = \frac{P(\vec{x} | \vec{\theta}) P(\vec{\theta})}{P(\vec{x})}
   \label{eq:bayesTheorem}
\end{equation}

\noindent where $\vec{x}$ and $\vec{\theta}$ correspond to a collection of data and model parameters, respectively.
The prior distribution, $P(\vec{\theta})$, explicitly denotes the choice of possible model parameters.
The likelihood, $P(\vec{x} | \vec{\theta})$, characterizes how well the model describes the data for a given set of parameters, including terms to account for the correlations between sources of uncertainties.
When these two terms are appropriately combined, we can extract the posterior distribution, $P(\vec{\theta} | \vec{x})$, which encodes the probability distribution of the parameters $\vec{\theta}$ given the data $\vec{x}$.
Note that the evidence, $P(\vec{x})$, is not required for these proceedings, and will be neglected.
Bayesian inference analysis, also known as a calibration, enables the investigation of broad questions of consistency between model and data, including the identification of regions of tension and areas for model improvements.

There are three key components of a Bayesian inference analysis: 1) the model selection and parametrization, 2) the data included in the calibration, 3) and the analysis itself.
Each of these components have a meaningful impact on the extracted posterior distributions, and should be carefully considered when evaluating a reported extraction.

The model selection is a choice of the analysis.
For hadronic collisions, Monte Carlo generators almost always required to explore the model parameter space, because of the nature of the observables.
Model parameters control the dynamics of parton-medium interactions through a chosen parametrization.
Energy loss in QCD matter is characterized by the jet transport coefficient, $\hat{q}$, which quantifies the momentum accumulated transverse to the direction of propagation of the probe per unit length.
Results of hard-sector calibrations are commonly  reported as a function of $\qhatTcubed{}$, which removes the leading $T$ dependence. There is a possible dependence of $\qhat{}$ on the momentum of the parton energy $E$, medium temperature $T$, and parton virtuality $Q$.
However, the comparison of different $\qhat{}$ extractions is challenging due to different choices in posterior distributions, parametrizations, and data~\cite{Apolinario:2022vzg}.

Data selection is also a critical component of the calibration, since different observables have different sensitivity to various aspects of medium evolution.
In general, one should aim to include all available experimental measurements of a particular observable rather than selecting subsets.
Differential studies with subsets of observables can then be performed to gain insight into the trends.
An accurate specification of uncertainties and their correlations is essential for a meaningful calibration.
Experiments should aspire to report full correlation matrices between sources of uncertainties, although signed uncertainties are often simpler to characterize and achieve a similar purpose.

For the Bayesian inference analysis, choices of the functional form of the parametrization of $\hat{q}$, as well of the prior distribution, are similarly critical.
Analyses in heavy-ion collisions are computationally expensive, requiring millions of core-hours for calculations at selected parameter values.
Computational techniques, such as active learning, transfer learning, and Gaussian processes are employed to optimize and reduce the required computational time.
Further details on the calibration procedures are available in Refs.~\cite{Bernhard:2019bmu, JETSCAPE:2020shq, Nijs:2020roc, Ke:2020clc, Xie:2022fak, JETSCAPE:2021ehl, JETSCAPE:2024cqe}.

\section{Calibrations with hadron measurements}

\begin{figure}[tb]
    \centering
    \includegraphics[width=0.53\linewidth]{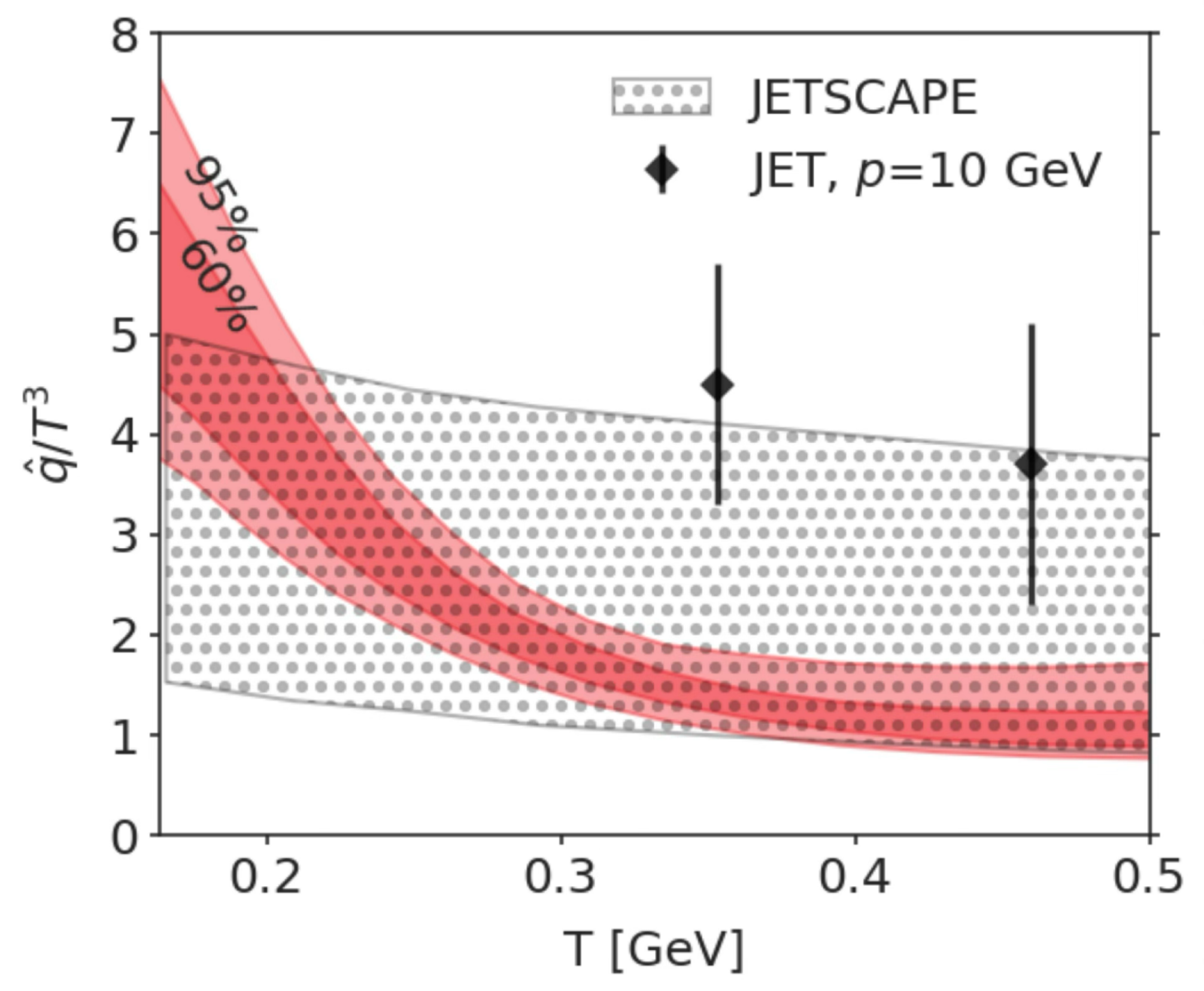}
    \caption{$\qhatTcubed{}$ extracted for a physics-inspired (dotted fill)~\cite{JETSCAPE:2021ehl} and an information field~\cite{Xie:2022ght,Xie:2022fak} (red) $\qhatTcubed{}$ parameterization calibrated on hadron yield modification data. Figure from Ref.~\cite{Xie:2022fak}.}
    \label{fig:hadronQHatComparison}
\end{figure}

Several recent Bayesian inference analyses have been performed using hadron yield modification measurements.
These include a proof-of-concept hard-sector calibration using a selection of hadron $\RAA{}$ measurements by the JETSCAPE Collaboration~\cite{JETSCAPE:2021ehl}, and a comprehensive calibration using all available inclusive hadron, dihadron, and $\gamma$-hadron yield modification measurements available at the time of publication~\cite{Xie:2022ght,Xie:2022fak}.
These analyses use data measured at both RHIC and LHC.
In addition to the differences in data selection, the analyses used different classes of $\qhat{}$ parametrizations.
The JETSCAPE analysis used a multistage approach using \texttt{MATTER} and \texttt{LBT} in the JETSCAPE framework~\cite{Putschke:2019yrg}, with a $\qhat{}$ parametrization motivated by the hard-thermal loop (HTL) formalism.
The analysis in Refs.~\cite{Xie:2022ght,Xie:2022fak} utilized an NLO parton model and higher twist formalism, with a $\qhat{}$ parametrization based on a flexible class of random functions known as an information field (IF), allowing the posterior distribution to take on a broad set of functional dependencies.
From here, the former analysis will be referred to as ``JETSCAPE hadron'', while the latter will be referred to as ``information field''.

Figure~\ref{fig:hadronQHatComparison} shows a comparison of the $\qhatTcubed{}$ posterior distributions extracted as a function of medium temperature for a quark with $E$ = 100 GeV propagating through the medium.
The distributions from the two analyses are also compared to the $\qhatTcubed{}$ values extracted from RHIC and LHC data by the JET collaboration~\cite{JET:2013cls}.
Despite the difference in data selection, their posterior distributions are consistent over much of the $T$ range, with the $\qhatTcubed{}$ distribution for the IF parametrization rapidly increasing at low $T$.
The 90\% credible interval (CI) of the IF posterior is consistently smaller than the JETSCAPE hadron posterior.

\begin{figure}[tb]
    \centering
    \includegraphics[width=0.49\linewidth]{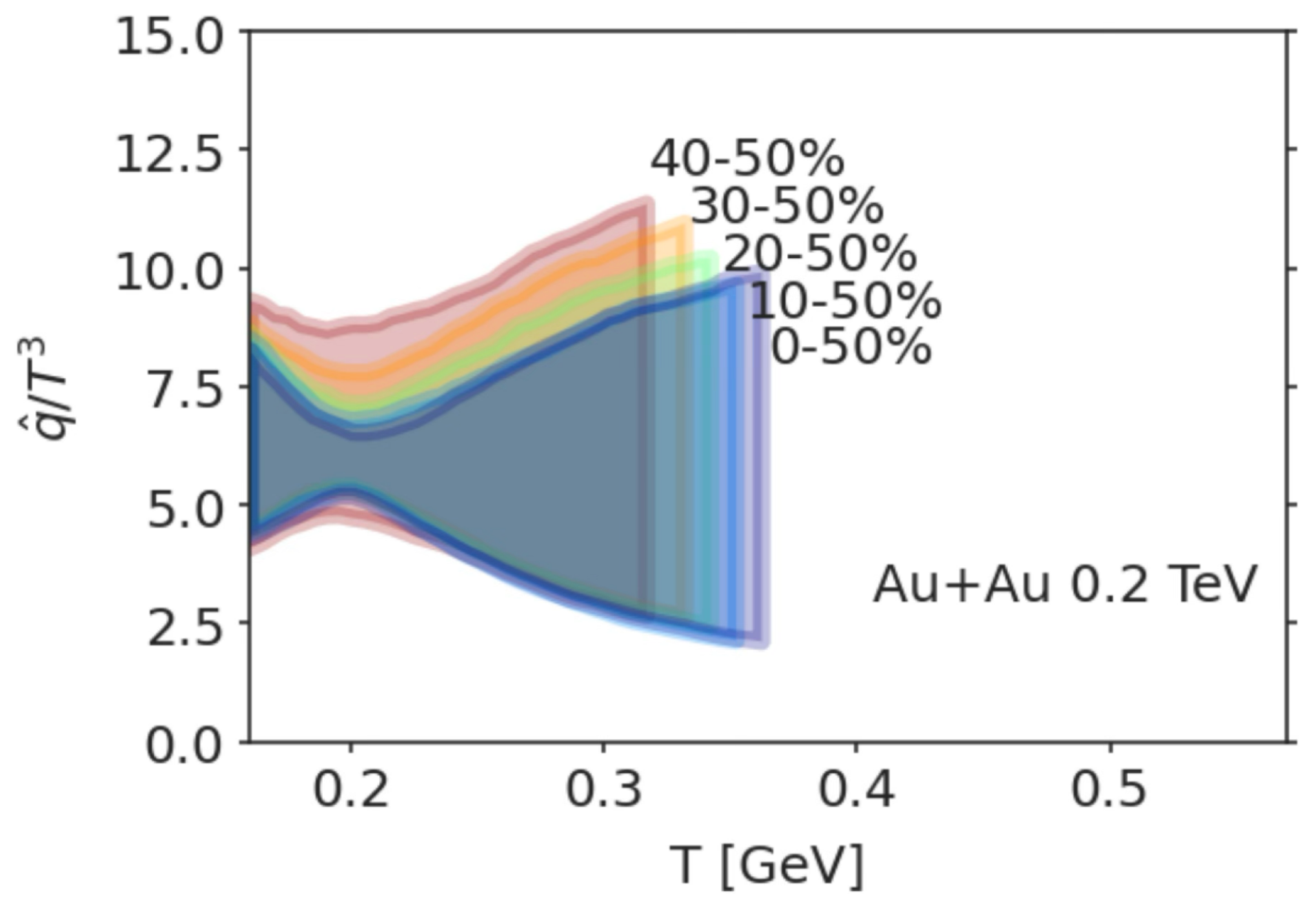}
    \includegraphics[width=0.49\linewidth]{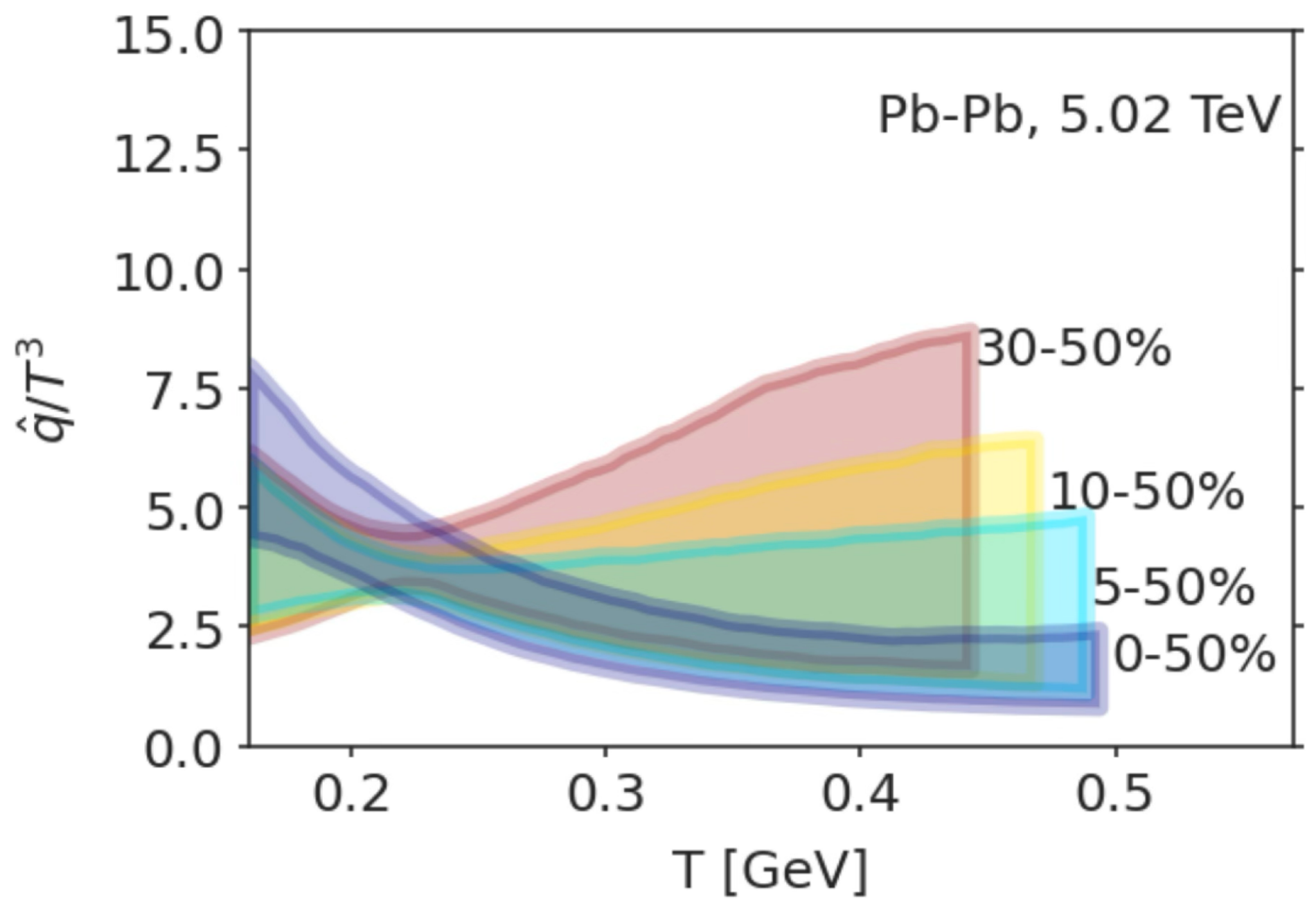}
    \caption{$\qhatTcubed{}$ extracted from differential selections of RHIC data measured at $\sqrtsNN{}$ = 200 GeV (left) and LHC data measured at $\sqrtsNN{}$ = 5.02 TeV (right) as a function of centrality selection. All calibrations are consistent at fixed $\sqrtsNN$, with improved constraints for more data. Figures from Ref.~\cite{Xie:2022fak}.}
    \label{fig:hadronRHICvsLHCQHat}
\end{figure}

These differences in the $\qhatTcubed{}$ posterior are driven by the model selection, data selection, and parametrization.
Although these differences cannot be fully disentangled, the low $T$ behavior provides insight into the impact of the parametrization choice.
This behavior is due to the lack of energy loss below the transition temperature, $T_{\text{c}}$, which is then compensated by increased energy loss above this transition.
This effect applies to both analyses, but only the IF parametrization is flexible enough to fully capture this dependence, leading to the rapid rise at low $T$.
However, in exchange for this flexibility, the parameters of the IF analysis lack a clear physics interpretation compared to the JETSCAPE hadron analysis.

The IF parameterization also enables investigations of the impact of differential data selections.
Figure~\ref{fig:hadronRHICvsLHCQHat} shows the $\qhatTcubed{}$ posterior distributions calibrated using varied centrality selections of data measured at RHIC and the LHC.
For the same collision energy, the calibrations performed using different centrality intervals are consistent within the 90\% CI.
As additional data are added by expanding the centrality range, the posterior distribution becomes more constrained at fixed $T$, indicating that the model description of the data as a function of centrality is consistent, and therefore more data provide stronger model constraints.
Since more central collisions reach higher temperatures, calibrations using more central data are able to constrain the posterior to larger $T$.
In contrast, less flexible parametrizations may introduce unphysical correlations between low and high $T$.

\section{Soft-hard calibrations}

\begin{figure}
    \centering
    \includegraphics[width=0.85\linewidth]{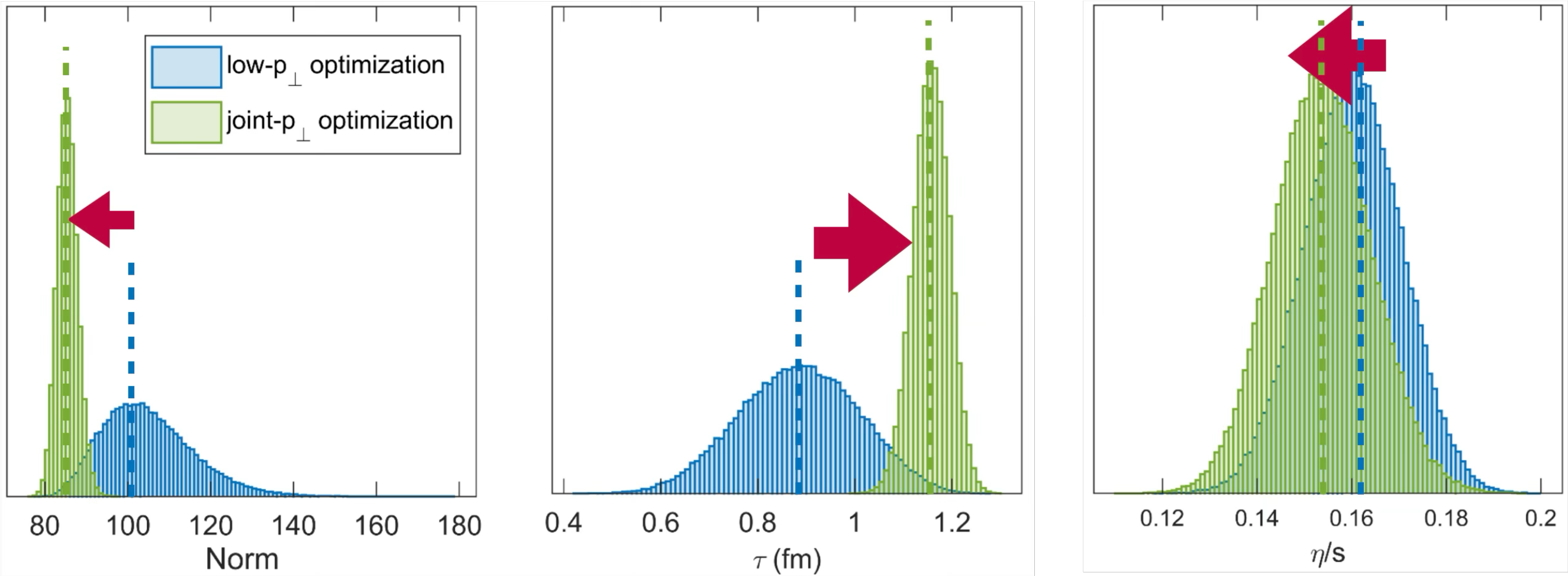}
    \caption{Parameter posterior distributions from the calibration of a subset of the soft-sector parameters in the DREENA-A model for low-$\pT{}$ (blue) and $\pT{}$-inclusive observables (green). As indicated by the red arrows, the most probable values of the distributions change with the $\pT{}$-inclusive data, although the distributions are consistent within their credible intervals. Figures are adapted from Ref.~\cite{Djordjevic:2024hp}.}
    \label{fig:dreenaHardSoftCorrelation}
\end{figure}

Although soft and hard sector calibrations have been treated separately to make calculations computationally tractable, there is growing interest in understanding how correlations between soft and hard observables impact the extracted model parameters.
An initial step in this direction was presented at the conference, performing a calibration of soft-sector parameters using soft and hard sector observables, including measurements of inclusive hadron and $D$-meson $\RAA{}$ and $v_{2}$.
Using the DREENA-A model and the parametrization described in Ref.~\cite{Karmakar:2023ity}, a subset of the soft-sector parameters were calibrated for low-$\pT{}$ observables only and for $\pT{}$-inclusive observables~\cite{Djordjevic:2024hp}.
The parameter posterior distributions of this calibration are shown in Fig.~\ref{fig:dreenaHardSoftCorrelation}.
Although the parameters are consistent between two calibrations, the most probable values are shifted when more data are included, indicating sensitivity of soft-sector parameters to the inclusion of high-$\pT{}$ observables.
These additional data also lead to more significant parameter constraints as compared to the low-$\pT{}$ observables alone.
Joint soft-hard calibrations are an important avenue of exploration in future Bayesian inference analyses as they become more computationally achievable.

\section{Calibrations with inclusive hadron and jet measurements}

\begin{figure}[tb]
    \centering
    \includegraphics[width=0.6\linewidth]{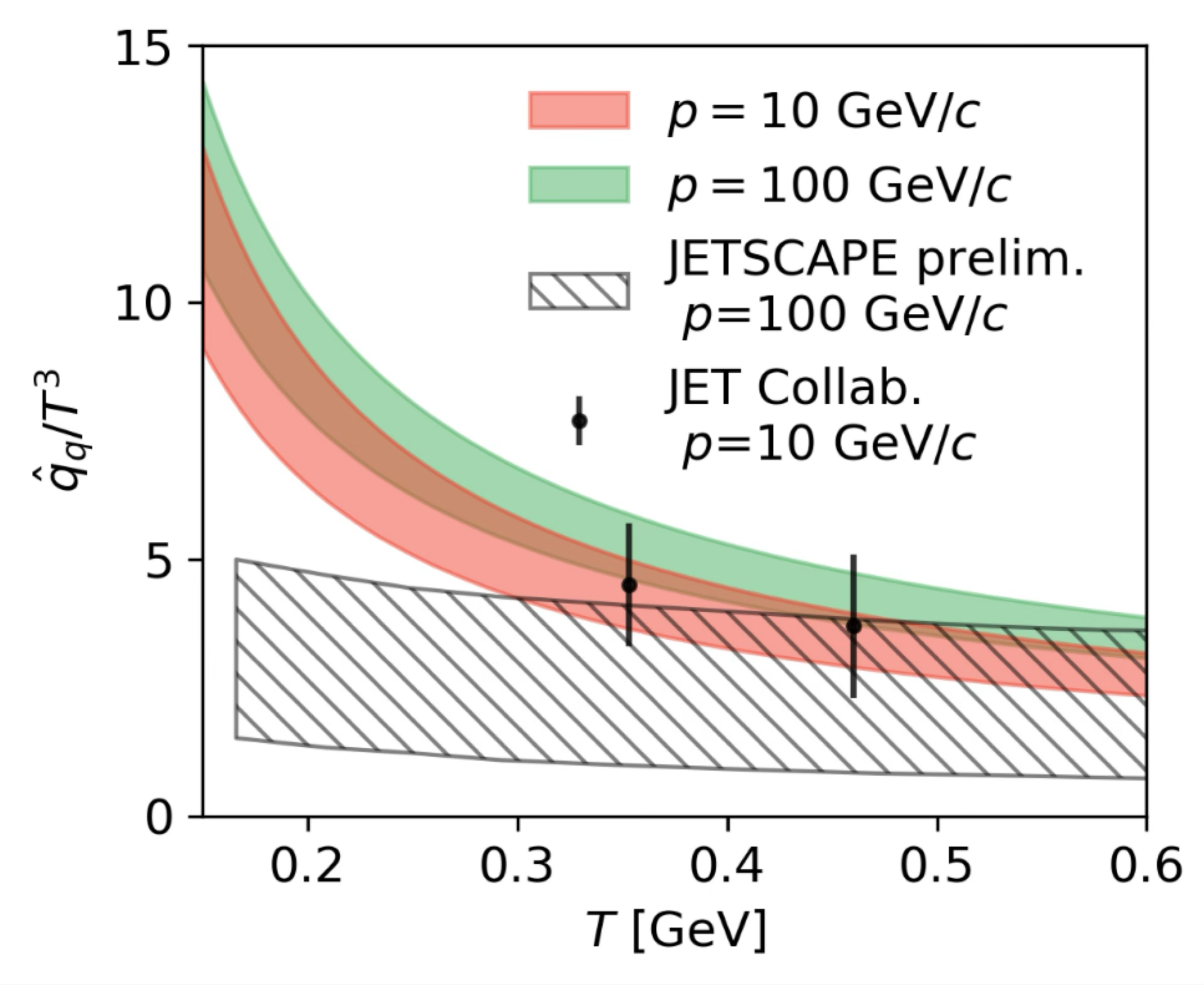}
    \caption{$\qhatTcubed{}$ posterior distributions as a function of $T$ extracted by the calibration of the LIDO model to selected hadron and jet measurements. $\qhatTcubed{}$ is shown for $E$ = 10 GeV (red) and $E$ = 100 GeV (green). Figure from Ref.~\cite{Ke:2020clc}.}
    \label{fig:qhatLIDO}
\end{figure}

A natural next step in calibrating the hard sector is to include reconstructed jet measurements alongside hadron observables.
The first Bayesian analysis to take this step was the calibration of the \texttt{LIDO} model to a selection of the most central inclusive hadron, inclusive jet, and $D$-meson yield modification measurements~\cite{Ke:2020clc}.
The $\qhatTcubed{}$ posterior distribution as a function of $T$ for two different quark energies $E$, is shown in Fig.~\ref{fig:qhatLIDO}.
This calibration extracted a constrained posterior distribution, which indicates a consistent description of both inclusive hadron and jet measurements.
The posterior rises more rapidly than in the JETSCAPE hadron calibration presented above, although the significant difference in the model and data selections prevents drawing strong conclusions.
These calibrated parameters were then used to predict additional observables not included in the Bayesian inference analysis.
Many observables are well described by these parameters, although some tension remains in others, such as the $R$ dependence of the jet $\RAA{}$, indicating regions for model improvements.

\begin{figure}[tb]
    \centering
    \includegraphics[width=0.9\linewidth]{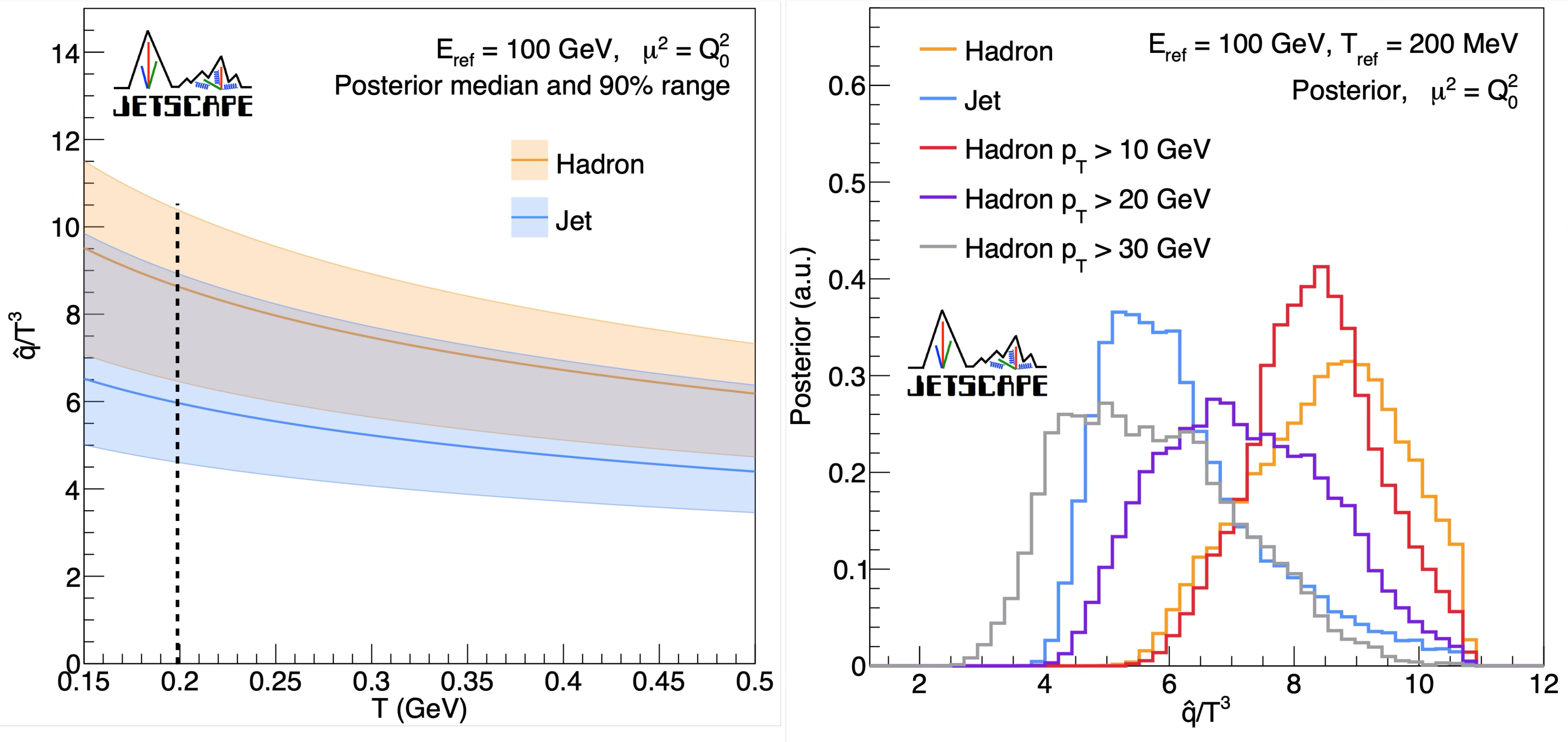}
    \caption{(left) $\qhatTcubed{}$ posterior distribution extracted for hadron-only and jet-only observables as a function of $T$. (Right) $\qhatTcubed{}$ posterior distribution extracted for jet-only observables (blue), hadron-only observables (orange), and selections of hadron data for $\pT{}$ $> 10$ \GeVc{} (red), $> 20$ \GeVc{} (purple), and $> 30$ \GeVc{} (gray). All curves are evaluated at $E$ = 100 GeV. The vertical dashed line on the left corresponds to $T$ = 200 MeV, which is where the distributions on the right are evaluated. Figures are adapted from Ref.~\cite{JETSCAPE:2024cqe}.}
    \label{fig:qhatJETSCAPE}
\end{figure}

The JETSCAPE Collaboration performed a comprehensive Bayesian inference analysis including all inclusive hadron and jet yield modification measurements available at the time of the analysis (February 2022).
JETSCAPE used a multistage simulation approach, transitioning from \texttt{MATTER} at high virtuality to \texttt{LBT} at low virtuality. The HTL-inspired $\qhat{}$ parametrization includes a reduction of interactions with the medium at high virtuality to account for coherence effects~\cite{JETSCAPE:2024cqe}.
The full $\qhatTcubed{}$ posterior distribution is consistent with previous extractions when evaluated at comparable virtuality.
However, calibrations performed using hadron-only and jet-only observables reveal some tension, as seen in the $\qhatTcubed{}$ posterior distributions shown on the left side of Fig.~\ref{fig:qhatJETSCAPE}.
Although the distributions are compatible within the 90\% CI, the jet-only and hadron-only posterior distributions bracket the combined calibration, with the jet-only (hadron-only) calibration preferring a lower (higher) $\qhatTcubed{}$.

If $\qhat{}$ is a universal property of the medium, then its extracted value should be independent of the observable.
These differences were therefore further explored via calibrations using selections of observables, including minimum and maximum $\pT{}$ requirements, as well as centrality selections.
While the maximum $\pT{}$ and centrality analyses did not explain the trend, the minimum hadron $\pT{}$ requirement provided deeper insight.
One-dimensional projections of the $\qhatTcubed{}$ posterior distribution for hadron $\pT{} > 10$--$30$ \GeVc{} at $E$ = 100 GeV and $T$ = 200 MeV are shown on the right of Figure~\ref{fig:qhatJETSCAPE}, and are also compared to the hadron-only and jet-only calibrations.
The hadron $\pT{}$-selected calibrations interpolate between the $\pT{}$-integrated hadron-only posterior and the jet-only posterior, with higher $\pT{}$ requirements leading to results which are more consistent with the jet-only calibration.
This discrepancy is driven by the low $\pT{}$ hadron measurements, which are the most precise experimentally and are therefore the most constraining in the Bayesian inference analysis.
While the theoretical model is expected to be most uncertain in this region, the current analysis does not incorporate theoretical uncertainties, and thus this effect is not taken into account.
When the low $\pT{}$ data are excluded, the model is better controlled and less constrained, and the calibration is more consistent.
This discrepancy points to the need to improve model descriptions at low $\pT{}$, as well as the importance of explicitly including theory uncertainties in Bayesian inference analyses.

Such theoretical uncertainties can account for differences between the model and the data not only due to model choices, but also due to missing physics, such as the lack of nuclear shadowing.
However, due to the multiscale sensitivity of these analyses, there is not yet a clear consensus on which issues are most pressing to address.
Some possibilities include moving beyond leading order in $\qhat{}$, refocusing on controllable vacuum parameters, or splitting processes into perturbative and non-perturbative sectors.

\section{Outlook}

Bayesian inference is an essential tool for characterizing and understanding the quark--gluon plasma. 
Calibrations performed using multiple observable provide deep insights into the dynamics and structure of the medium, far beyond what is achievable by comparisons to single observables.
In these proceedings, the components and tools necessary to evaluate Bayesian inference analyses were discussed.
Substantial progress has been made in recent years on such analyses in the hard sector, rapidly expanding from a proof-of-concept calibration using a limited set of hadron observables to multiple comprehensive analyses utilizing the world's data on hadron and jet yield modification.

There are several promising avenues for investigation in the coming years, spanning all three main components of Bayesian inference.
With these comprehensive calibrations in hand, studies of observable sensitivity and experimental design will be performed to determine which observables will have the greatest impact on calibrations.
These studies will provide guidance for which observables are most critical to measure with high precision.
Based on the calibrations so far, it appears likely that new high-precision high-$\pT{}$ hadron $\RAA{}$ measurements at RHIC and high-precision jet substructure measurements will be highly impactful.
The posterior distributions from current analyses can also be used to predict new observables not included in the calibration, to study the limits of model applicability.

Key developments in the near future include adding new observables, improving computational efficiency, and performing multi-model calibrations.
New Bayesian inference analyses will systematically incorporate additional observables to characterize their sensitivity to QGP parameters.
JETSCAPE has presented preliminary results on the inclusion of jet substructure observables, demonstrating stronger constraints on $\qhatTcubed{}$ due to the complementary information carried by those observables~\cite{JETSCAPE:2024ofm}.

At the level of Bayesian analyses, investigations into the use of informative (i.e., non-uniform) priors to improve constraining power are ongoing, taking full advantage of the knowledge gained from existing comprehensive calibrations.
Reporting signed experimental uncertainties, as well as fully treating theoretical uncertainties, will improve the robustness of the calibrations.
Novel computational methods, notably those based on machine learning, will play a critical role by reducing computational costs.
Such efficiency improvements enable new analyses, such as a full soft-hard calibration.

A key goal of next-generation analyses should be to compare multiple models under equivalent conditions.
Ideally, this should include identical treatment of soft-sector evolution, as well as comparable $\qhat{}$ parametrizations.
Such comparisons will usher in a new era of model discrimination, utilizing the full wealth of RHIC and LHC data to constrain the microscopic physics of the QGP.

\bibliography{main.bib}

\end{document}